*Review*

# Broadband Transformation Optics Devices

**Vera N. Smolyaninova** [1,]*, **Igor I. Smolyaninov** [2], **Alexander V. Kildishev** [3] and **Vladimir M. Shalaev** [3]

[1] Department of Physics, Astronomy & Geosciences, Towson University, 8000 York Road, Towson, Maryland 21252-0001, USA

[2] Electrical and Computer Engineering Department, University of Maryland, College Park, Maryland 20742, USA; E-Mail: smoly@umd.edy (I.I.S.)

[3] Electrical School of Electrical and Computer Engineering and Birck Nanotechnology Center, Purdue University, West Lafayette, Indiana 47907, USA; E-Mails: kildishev@purdue.edu (A.V.K.); shalaev@purdue.edu (V.M.S.)

* Author to whom correspondence should be addressed; E-Mail: vsmolyaninova@towson.edu;

**Abstract:** Recently we have suggested that two-dimensional broadband transformation optics devices based on metamaterial designs may be built using tapered waveguides. Here we review application of this principle to broadband electromagnetic cloaking, trapped rainbow, and novel microscopy devices.

## 1. Introduction

Current interest in electromagnetic metamaterials has been motivated by recent work on superlenses, cloaking and transformation optics [1-3]. This interest has been followed by considerable efforts aimed at introduction of metamaterial structures that could be realized experimentally. Unfortunately, it appears difficult to develop metamaterials with low-loss, broadband performance. The difficulties are especially severe in the visible frequency range where good magnetic performance is limited. On the other hand, very recently we have demonstrated that many transformation optics and metamaterial-based devices requiring anisotropic dielectric permittivity and magnetic permeability could be emulated by specially designed tapered waveguides [4]. This approach leads to low-loss, broadband performance in the visible frequency range, which is difficult to achieve by other means. We have applied this technique to broadband electromagnetic cloaking in the visible range [4], first experimental demonstration of the "trapped rainbow" [5], and experimental realization of the Maxwell fisheye and inverted Eaton microlenses [6], which were suggested to act as superb imaging devices in the absence of negative refraction [7]. Realization of these devices using electromagnetic metamaterials would require sophisticated nanofabrication techniques. In contrast, our approach leads to a much simpler design, which requires readily available dielectric materials. In this article we will describe in detail our recent experiments on the broadband transformation optics devices based on this principle.

## 2. Broadband Cloaking in the Visible Frequency Range

Until recently, the experimental realizations of cloaking-like behavior were limited to very small sizes, on the scale of a few wavelengths only [8,9]. In such cases, light scattering and shadow produced by an uncloaked object of the same size would not be strong anyway. Let us demonstrate that metamaterial devices, such as large cloaks requiring anisotropic dielectric permittivity and magnetic permeability can be emulated by specially designed tapered waveguides. This approach leads to low-loss, broadband performance in the visible frequency range, which is difficult to achieve by other means. We apply this technique to electromagnetic cloaking and demonstrate broadband, two-dimensional electromagnetic cloaking in the visible frequency range on a scale roughly 100 times larger than that of the incident wavelength.

As a starting point, we show that the transformation optics approach allows us to map a planar region of space filled with an inhomogeneous, anisotropic metamaterial into an equivalent region of empty space with curvilinear boundaries (a tapered waveguide). We begin with Maxwell's curl equations

$$\tilde{\nabla} \times \tilde{H} = -i\omega\tilde{\varepsilon}\tilde{E}, \quad \tilde{\nabla} \times \tilde{E} = i\omega\tilde{\mu}\tilde{H} \tag{1}$$

for vector fields $\tilde{E} = \sum \tilde{e}_i \tilde{x}_i$ and $\tilde{H} = \sum \tilde{h}_i \tilde{x}_i$ in an orthogonal curvilinear system with the unit vectors $\tilde{x}_i$. The components of the actual physical fields $E = \sum e_i \tilde{x}_i$ and $H = \sum h_i \tilde{x}_i$ are connected with the fields in the material coordinates through $e_i = \tilde{e}_i s_i^{-1}$ and $h_i = \tilde{h}_i s_i^{-1}$ via the metric coefficients $s_i$. The tensors $\tilde{\varepsilon}$ and $\tilde{\mu}$ are given by $t\varepsilon$ and $t\mu$, with **t** being $t = s_1 s_2 s_3 diag(s_1^{-2}, s_2^{-2}, s_3^{-2})$. For comparing axisymmetric cloaking and imaging systems, it is convenient to match a given axisymmetric material domain to an equivalent inhomogeneous axisymmetric material distribution between two planes in the cylinder coordinate system, as shown in Figure 1a. In this case the metric coefficients are $s_\rho = s_z = 1$, $s_\varphi = \rho$. Therefore, $\rho e_\phi = \tilde{e}_\phi$, $\rho h_\phi = \tilde{h}_\phi^2$, and $e_i = \tilde{e}_i$, $h_i = \tilde{h}_i$ for i = ρ, z. Thus, any rotational coordinate system $(\tilde{\rho}, \tilde{\phi}, \tilde{z})$ converted into the cylindrical format

$$\tilde{\nabla} \times \tilde{H} = -i\omega diag(\tilde{\rho}, \tilde{\rho}^{-1}, \tilde{\rho})\tilde{\varepsilon}\tilde{E}, \quad \tilde{\nabla} \times \tilde{E} = i\omega diag(\tilde{\rho}, \tilde{\rho}^{-1}, \tilde{\rho})\tilde{\mu}\tilde{H} \tag{2}$$

is equivalent to a cylinder coordinate system with material tensors $\tilde{\varepsilon}$ and $\tilde{\mu}$.

**Figure 1.** A space between a spherical surface and a planar surface (**a**) mapped onto a layer with planar boundaries (**b**). (**c**) Distribution of the radial (top), azimuthal (middle), and axial (or vertical) diagonal components of permittivity and permeability in the equivalent planar waveguide. Dashed lines show the same components in the waveguide with a radius-dependent refractive index. (**d**) Normalized profile of the optimal waveguide shape plotted for a cloak radius of $b_0 = 172$ μm. The shape of the optimal waveguide may be approximated by a spherical surface placed on top of a flat surface, as shown by the dashed line.

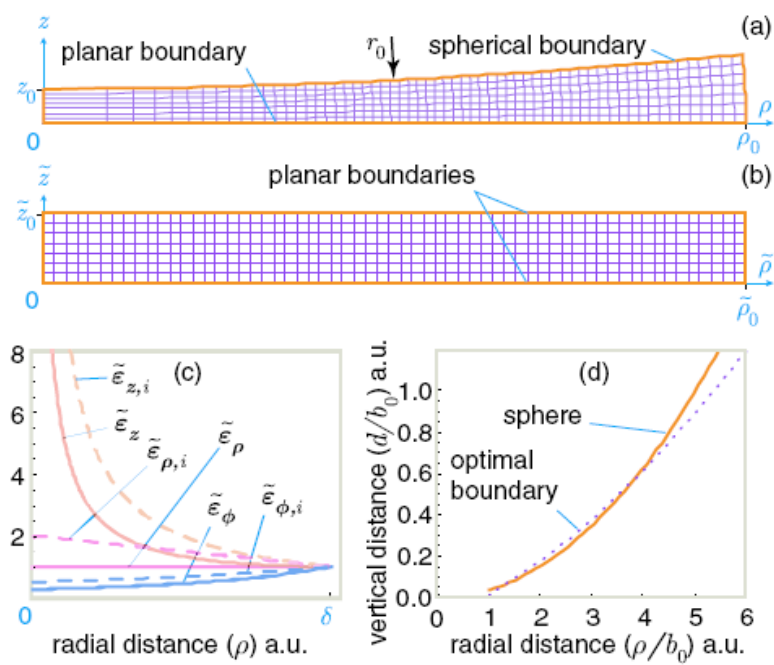

Let us now consider an interesting application of the formalism above, which will lead us to an experimental demonstration of electromagnetic cloaking. We map an axisymmetric space between two spherical surfaces onto a space between two planes. The parametric description,

$$s^2 = (z - z_0)^2 + \rho^2, \quad \rho = \frac{s\sqrt{(2s - \tilde{\rho})\tilde{\rho}}}{s - \tilde{\rho} + \sqrt{s^2 + \tilde{z}^2}}, \quad \rho = \frac{s\,\tilde{z}}{s - \tilde{\rho} + \sqrt{s^2 + \tilde{z}^2}} \tag{3}$$

provides this mapping (see Figure 1b, note that $\varphi = \tilde{\phi}$). The transformation optics technique gives the diagonal components

$$\tilde{\varepsilon}_{\tilde{\rho}} = \frac{s(2s - \tilde{\rho})}{a(a + s - \tilde{\rho})}, \quad \tilde{\varepsilon}_{\phi} = \frac{s^3}{a(a + s - \tilde{\rho})(2s - \tilde{\rho})}, \quad \tilde{\varepsilon}_{\tilde{z}} = \frac{as}{\tilde{\rho}(a + s - \tilde{\rho})} \tag{4}$$

of $\tilde{\mu} = \tilde{\varepsilon}$ tensors, distributed in an equivalent layer between two planes (where $a = \sqrt{s^2 + z^2}$ ). Analysis of Equations (4) and Figure 1(c) indicates that Equations (4) can be approximated with

$$\tilde{\varepsilon_{\tilde{\rho}}} = \tilde{\mu_{\tilde{\rho}}} \approx 1, \quad \tilde{\varepsilon_{\tilde{\phi}}} = \tilde{\mu_{\tilde{\phi}}} \approx \frac{s^2}{(2s - \tilde{\rho})^2}, \quad \tilde{\varepsilon_{\tilde{z}}} = \tilde{\mu_{\tilde{z}}} \approx \frac{s^2}{\tilde{\rho}(2s - \tilde{\rho})} \tag{5}$$

Note that the requirement for an ideal cloak should be written as [8]

$$\tilde{\varepsilon_{\tilde{\rho}}} = \tilde{\mu_{\tilde{\rho}}} = \rho \tilde{\rho}' / \tilde{\rho}, \quad \tilde{\varepsilon_{\tilde{\phi}}} = \tilde{\mu_{\tilde{\phi}}} = \tilde{\varepsilon_{\tilde{\rho}}}^{-1}, \quad \tilde{\varepsilon_{\tilde{z}}} = \tilde{\mu_{\tilde{z}}} = \rho / (\tilde{\rho}' \tilde{\rho}), \tag{6}$$

where $\tilde{\rho} = \tilde{\rho}(\rho)$ is a radial mapping function and $\tilde{\rho}' = d\tilde{\rho} / d\rho$ . These requirements can be met if the refractive index ($n = \varepsilon^{1/2}$ ) inside the gap between the sphere and the plane is chosen to be a simple radius-dependent function $n = \sqrt{(2s - \tilde{\rho}) / s}$ . In such a case we obtain

$$\tilde{\varepsilon_{\tilde{\rho}}} \approx \frac{2s - \tilde{\rho}}{s}, \quad \tilde{\varepsilon_{\tilde{\phi}}} = \frac{s}{2s - \tilde{\rho}}, \quad \tilde{\varepsilon_{\tilde{z}}} = \frac{s}{\tilde{\rho}} \tag{7}$$

Note that the scale $s$ is chosen to avoid singularities: $s > \max(\tilde{\rho})$, and the filling substance has an isotropic effective refractive index ranging from 2 to 1 for $\rho = [0, s]$. Equation (7) represents the *invisible body*, *i.e.*, a self-cloaking arrangement. It is important to note that filling an initial domain between a rotationally symmetric curvilinear boundaries, for example, with anisotropic dielectric allows for independent control over the effective magnetic and electric properties in the equivalent right-cylinder domain.

In the semi-classical ray-optics approximation, the cloaking geometry may be simplified further for a family of rays with similar parameters. Our starting point is the semi-classical 2D cloaking Hamiltonian (dispersion law) introduced in [10]:

$$\frac{\omega^2}{c^2} = k_r^2 + \frac{k_\phi^2}{(r-b)^2} = k_r^2 + \frac{k_\phi^2}{r^2} + k_\phi^2 \frac{b(2r-b)}{(r-b)^2 r^2} \tag{8}$$

For such a cylindrically symmetric Hamiltonian, the rays of light would flow without scattering around a cloaked region of radius $b$. Our aim is to produce the Hamiltonian (8) in an optical waveguide (Figure 2). Let us allow the thickness $d$ of the waveguide in the $z$-direction to change adiabatically with radius $r$. The top and bottom surfaces of the waveguide are coated with metal. The dispersion law of light in such a waveguide is

$$\frac{\omega^2}{c^2} = k_r^2 + \frac{k_\phi^2}{r^2} + \frac{\pi^2 m^2}{d(r)^2} \tag{9}$$

where $m$ is the transverse mode number. A photon launched into the $m$-th mode of the waveguide stays in this mode as long as $d$ changes adiabatically [11]. Since the angular momentum of the photon $k_\phi = L$ is conserved, for each combination of $m$ and $L$ the cloaking Hamiltonian (8) can be emulated precisely by an adiabatically changing $d(r)$. Comparison of Equations (8) and (9) produces the following desired radial dependence of the waveguide thickness:

$$d = \frac{m\pi r^{3/2}\left(1 - \dfrac{b_{mL}}{r}\right)}{L\left(2b_{mL}\left(1 - \dfrac{b_{mL}}{r}\right)\right)^{1/2}} \qquad (10)$$

where $b_{mL}$ is the radius of the region that is "cloaked" for the photon launched into the *(m,L)* mode of the waveguide. The shape of such a waveguide is presented in Figure 1(d).

**Figure 2.** Tapered waveguide acting as an optical cloak (not to scale). (**a**) Cross-section sketch for the waveguide experiment. (**b**) 3D rendition of the experimental setup sketched in (a); in the figure one-quarter of the gold-coated lens is removed to show the interior details.

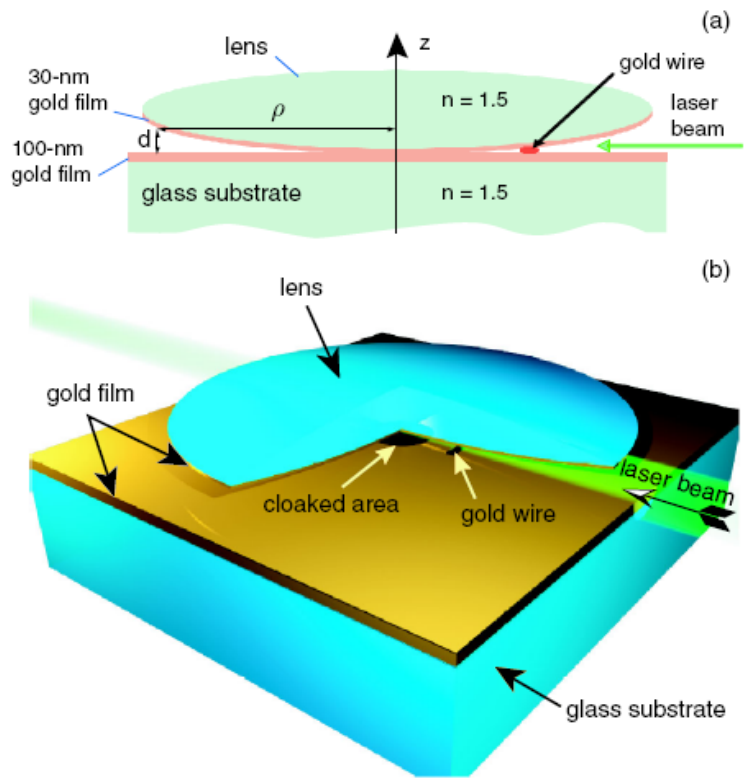

Thus, an electromagnetic cloaking experiment in a waveguide may be performed in a geometry that is identical to the classic geometry of the Newton rings observation, as shown in Figure 2. An aspherical lens shaped according to Equation (10) has to be used for the single *(m,L)* mode cloak to be ideal. This shape may be approximated by a spherical surface placed on top of a flat surface as shown by the dashed line in Figure 1 (d). Moreover, the waveguide geometry may be further improved to allow cloaking in a multi-mode waveguide geometry. The choice of $b_{mL} = b_o \dfrac{m^2}{L^2}$ leads to the same desired shape of the waveguide for all the *(m,L)* modes of a multi-mode waveguide in the leading order of $b_{mL}/r$ : $d = \pi\left(r^3/2b_0\right)^{1/2}$. This equation describes the best-shaped aspherical lens for the electromagnetic cloaking observation.

In our experiments, a 4.5-mm diameter double convex glass lens (lens focus 6 mm) was coated on one side with a 30-nm gold film. The lens was placed with the gold-coated side down on top of a flat glass slide coated with a 70-nm gold film. The air gap between these surfaces has been used as an adiabatically changing waveguide. The point of contact between two gold-coated surfaces is visible in Figure 3(a). The Newton rings appear around the point of contact upon illumination of the waveguide with white light from the top. The radius of the *m*-th ring is given by the expression $r_m = ((1/2+m)R\lambda)^{1/2}$ , where *R* is the lens radius. The central area around the point of contact appears bright since light reflected from the two gold coated surfaces has the same phase. Light from an argon ion laser was coupled to the waveguide via side illumination. Light propagation through the waveguide was imaged from the top using an optical microscope. Figure 3 show microscope images of the light propagation through the waveguide in an experiment in which a gold particle cut from a 50-μm diameter gold wire is placed inside the waveguide. A pronounced long shadow is cast by the particle inside the waveguide (Figure 3(e)). This is natural since the gold particle size is approximately equal to $100\lambda$ (note that the first dark Newton ring visible in Figures 3(a,b) has approximately the same size). Since the gold particle is located ~ 400 μm from the point of contact between the walls of the waveguide, the effective Hamiltonian around the gold particle differs strongly from the cloaking Hamiltonian of Equation (8). Figure 3 represents the results of our best effort to insert a 150 μm long 50 μm diameter gold particle inside the waveguide and orient it along the illumination direction. A few scratches visible in Figure 3 resulted from achieving this task.

While the gold particle casts a long and pronounced shadow, the area around the point of contact between the two gold-coated surfaces casts no shadow at all (Figure 4). This is an observation which would be extremely surprising in the absence of the theoretical description presented above. For the *m*-th mode of the waveguide shown in Figure 2, the cut-off radius is given by the same expression as that of the radius of the *m*-th Newton ring, which means that no photon launched into the waveguide can reach an area within the radius $r_0 = (R\lambda/2)^{1/2}$~ 30 μm from the point of contact between the two gold-coated surfaces. This is consistent with the fact that the area around the point of contact appears dark in Figure 4. Even though photons may couple to surface plasmons [12] near the cut-off point of the waveguide, the propagation length of the surface plasmons at 515 nm is only a few micrometers. Thus, the area around the point of contact about 50 μm in diameter is about as opaque for guided photons as the ~50-μm gold particle from Figure 3(c), which casts a pronounced shadow. Nevertheless, there appears to be no shadow behind the cut-off area of the waveguide (Figure 4). The observed cloaking behavior appears to be broadband (the images in Figure 4 were taken at 488-nm and 515-nm laser lines), which is consistent with the theory presented above.

**Figure 3.** (**a**) Microscope image of the waveguide without a particle (white light coming from the top). (**b**) Magnified Newton rings taken from the frame in (a). (**c**) Image of the waveguide with a gold particle (white light coming from the top). (**d**) Magnified Newton rings taken from the frame in (c). A long shadow cast by the gold particle upon coupling (via side illumination) 515-nm (**e**) and 488-nm (**g**) laser light into the waveguide. Magnified images of the rings for 515-nm (**f**) and 488-nm (**h**) light taken from frames in (e) and (g).

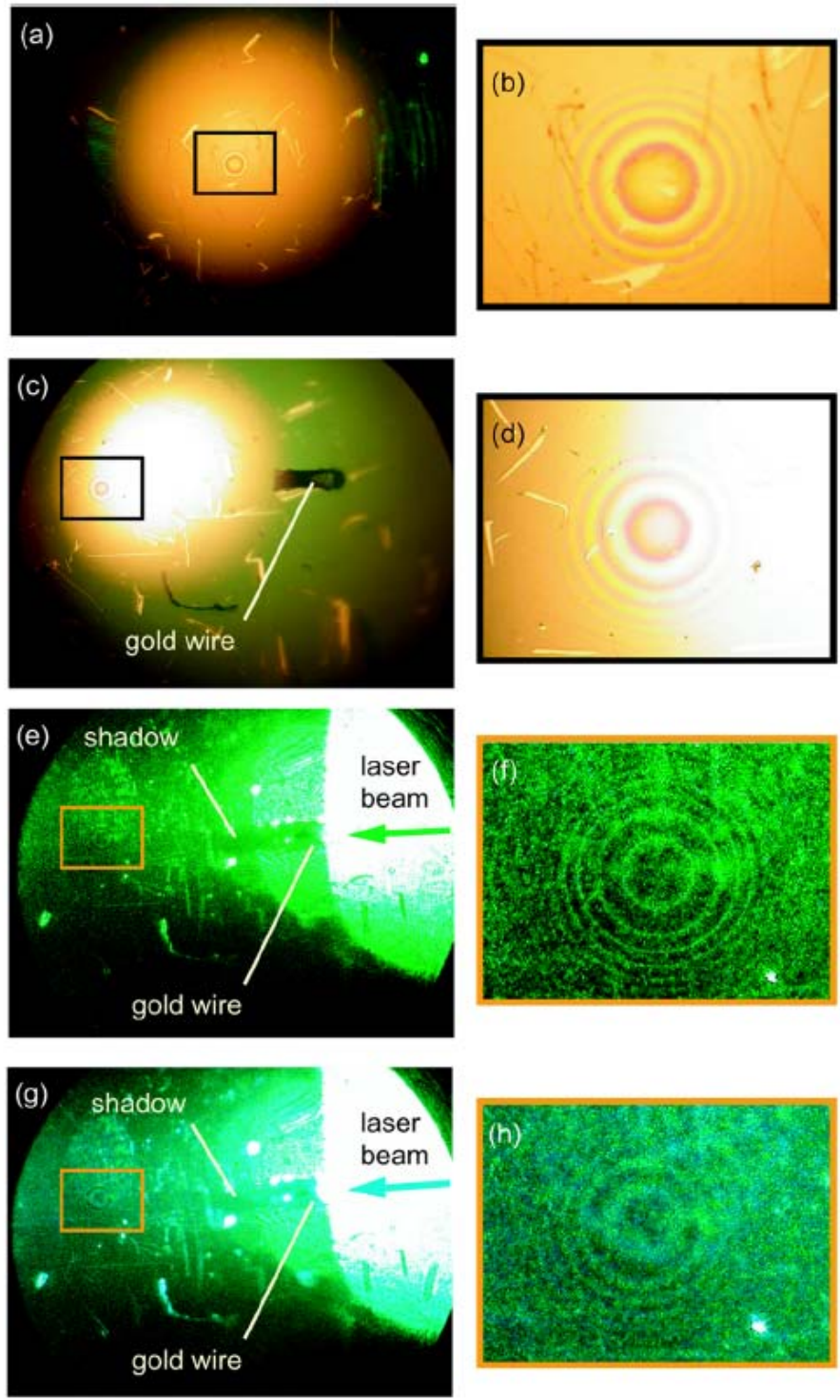

**Figure 4.** Magnified images of the rings for (**a**) 488-nm and (**b**) 515-nm laser illumination. In both cases, no wire is placed in the waveguide.

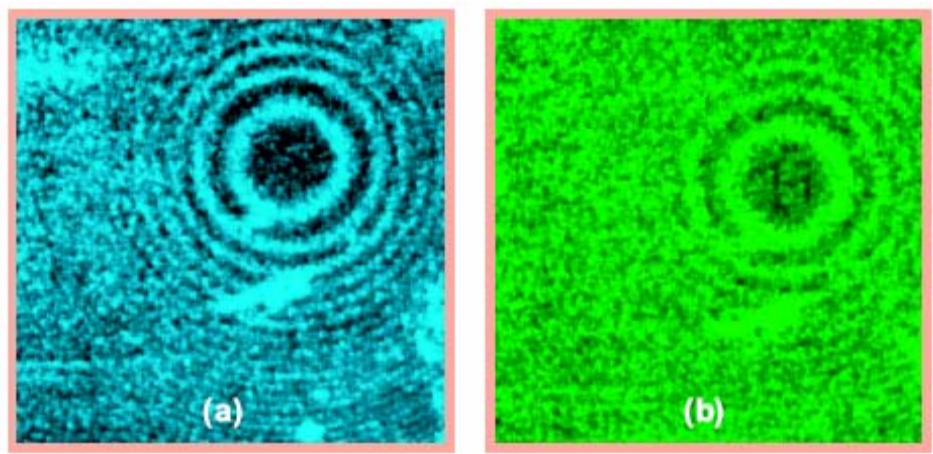

## 3. Experimental Observation of the Trapped Rainbow

The concept of a "trapped rainbow" has attracted considerable recent attention. According to various theoretical models, a specially designed metamaterial [13] or plasmonic [14,15] waveguide has the ability to slow down and stop light of different wavelengths at different spatial locations along the waveguide, which is extremely attractive for such applications as spectroscopy on a chip. In addition, being a special case of the slow light phenomenon [16], the trapped rainbow effect may be used in applications such as optical signal processing and enhanced light-matter interactions [17]. On the other hand, unlike the typical slow light schemes, the proposed theoretical trapped rainbow arrangements are extremely broadband, and can trap a true rainbow ranging from violet to red in the visible spectrum. Unfortunately, due to the necessity of complicated nanofabrication and the difficulty of producing broadband metamaterials, the trapped rainbow schemes had until recently remained in the theoretical domain only.

Very recently we have demonstrated an experimental realization of the broadband trapped rainbow effect which spans the 457–633 nm range of the visible spectrum [5]. Similar to our recent demonstration of broadband cloaking [4], the metamaterial properties necessary for device fabrication were emulated using an adiabatically tapered optical nano waveguide geometry. A 4.5-mm diameter double convex glass lens was coated on one side with a 30-nm gold film. The lens was placed with the gold-coated side down on top of a flat glass slide coated with a 70-nm gold film (Figure 5(a)). The air gap between these surfaces has been used as an adiabatically changing optical nano waveguide. The dispersion law of light in such a waveguide is described by Equation (9). Light from a multi-wavelength argon ion laser (operating at $\lambda$ = 457 nm, 465 nm, 476 nm, 488 nm and 514 nm) and 633-nm light from a He-Ne laser were coupled to the waveguide via side illumination. This multi-line illumination produced the appearance of white light illuminating the waveguide (Figure 5(b)). Light propagation through the nano waveguide was imaged from the top using an optical microscope (Figure 5(c)).

**Figure 5.** (**a**) Experimental geometry of the trapped rainbow experiment: a glass lens was coated on one side with a gold film. The lens was placed with the gold-coated side down on top of a flat glass slide also coated with a gold film. The air gap between these surfaces formed an adiabatically changing optical nano waveguide. (**b**) Photo of the trapped rainbow experiment: HeNe and Ar:Ion laser light is coupled into the waveguide. (**c**) Optical microscope image of the trapped rainbow.

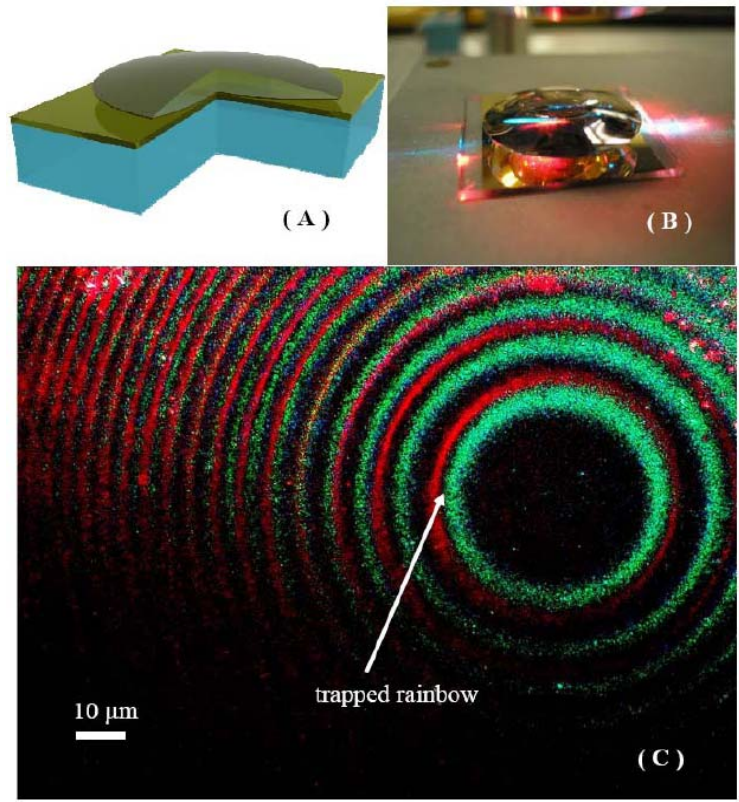

Since the waveguide width at the entrance point is large, the air gap waveguide starts as a multi-mode waveguide. Note that a photon launched into the *m*-th mode of the waveguide stays in this mode as long as *d* changes adiabatically [11]. In addition, the angular momentum of the photons $k_\phi = \rho k = L$ is conserved (where $\rho$ is the impact parameter defined with respect to the origin). Gradual tapering of the waveguide leads to mode number reduction: similar to our observation of broadband cloaking described above, only $L = 0$ modes may reach the vicinity of the point of contact between the gold-coated spherical and planar surfaces, and the group velocity of these modes

$$c_{gr} = c\sqrt{1 - \left(\frac{m\lambda}{2d}\right)^2} \tag{11}$$

tends to zero as *d* is reduced: the colored rings around the central circular dark area in Figure 5(c) each represent a location where the group velocity of the *m*-th waveguide mode becomes zero. These locations are defined by

$$r_l = \sqrt{(m + 1/2)R\lambda} \tag{12}$$

where R is the lens radius. Finally, the light in the waveguide is completely stopped at a distance

$$r = \sqrt{R\lambda / 2} \tag{13}$$

from the point of contact between the gold-coated surfaces, where the optical nano waveguide width reaches $d = \lambda/2\sim200$ nm range. The group velocity of the only remaining waveguide mode at this point is zero. This is consistent with the fact that the area around the point of contact appears dark in Figure 5(c). In this area the waveguide width falls below 200 nm down to zero. Since the stop radius depends on the light wavelength, different light colors stop at different locations inside the waveguide, which is quite obvious from Figure 5(c). Thus, the visible light rainbow has been stopped and "trapped." This observation constitutes an experimental demonstration of a broadband trapped rainbow effect in the visible frequency range. Unlike other recently proposed metamaterial-based schemes, our geometry is easily scalable to any spectral range of interest. While the group velocity of the trapped photons is exactly zero at $r = \sqrt{R\lambda / 2}$ (see Equation (11)), light cannot be "stored" indefinitely at these locations due to Joule losses in metal. In the best case scenario photons can be stored for no longer that 100–1000 periods. However, even this duration is enough to cause considerable enhancement of light-matter interaction in this geometry. Note that the light is stopped only for the waveguide mode, which has both the mode number $m = 0$ and the angular momentum number $L = 0$. Therefore, our current result does not contradict our observations of cloaking reported in ref. [4]. In the ray optics approximation this condition corresponds to the central ray hitting the cloak. As was noted in ref. [2], such a ray "does not know" which way to turn around the cloak.

The described experimental arrangement may be used in such important applications as spectroscopy on a chip. Figure 6 presents a comparison of the optical microscope images of the trapped rainbow effect from Figure 5(c) and the image obtained when only two laser wavelengths (514 nm and 633-nm) are used for illumination (shown at the top of Figure 6). Individual spectral lines separated by only a few micrometers appear to be well resolved in the latter image, which is evident from the cross section analysis presented in Figure 7. Based on the image cross section analysis, spectral resolution of the order of 40 nm has been obtained. Further improvement of spectral resolution may be achieved by using a gold-coated spherical surface with a smaller radius of curvature.

**Figure 6.** Comparison of the optical microscope images of the trapped rainbow effect from Figure 5(c) (right panel) and the image obtained when only two laser wavelengths (514 nm and 633-nm) are used for illumination (left panel).

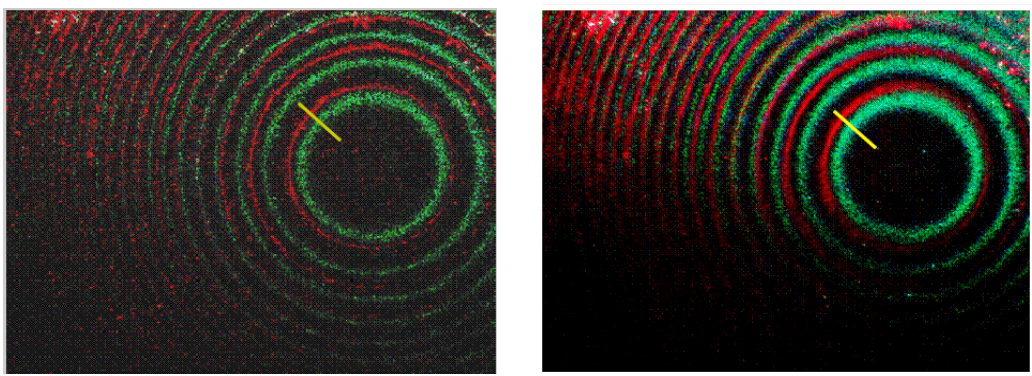

**Figure 7.** Cross sections of the optical microscope images along the yellow lines shown in Figure 6. Individual spectral lines are clearly resolved in the left plot obtained using 514 nm and 633 nm illumination. Multiple spectral lines are visible in the right plot, which is obtained under multiline illumination.

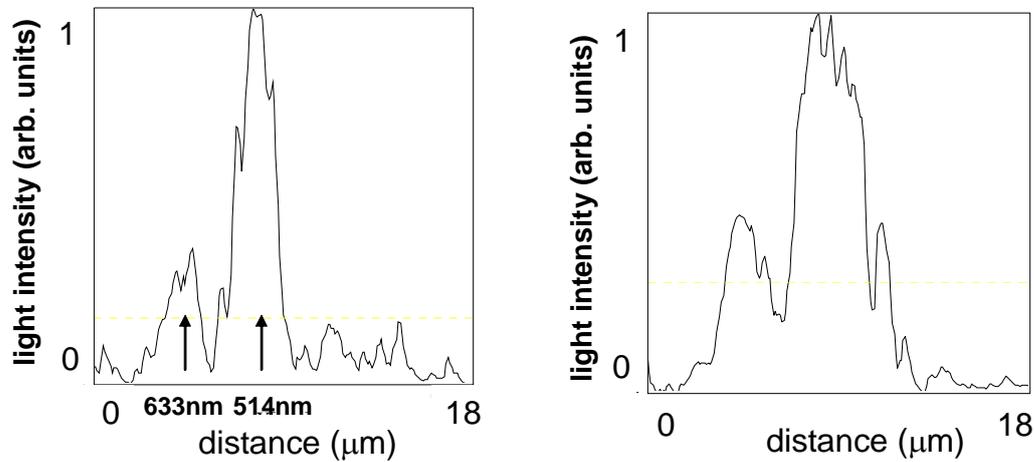

## 4. Maxwell Fisheye and Eaton Lenses Emulated by Microdroplets

Now we can apply a similar technique to experimental realization of the Maxwell fisheye and inverted Eaton microlenses, which were suggested to act as superb imaging devices even in the absence of negative refraction [7]. Realization of these microlenses using electromagnetic metamaterials would require sophisticated nanofabrication techniques. In contrast, our approach leads to a much simpler design, which involves two-dimensional (2D) imaging using a small liquid microdroplet.

We should note that despite strong experimental and theoretical evidence supporting superresolution imaging based on microlenses and microdroplets, imaging mechanisms involved are not well understood. Imaging by surface plasmon polaritons (SPP) [12] has been proposed as the main super-resolution mechanism in imaging experiments using glycerin microdroplets on gold film surface [18]. Resolution of the order of λ/8 has been observed in these experiments. On the other hand, magnification of near-field image components has been suggested in recent experiments with self-assembled plano-spherical nanolenses [19,20], which demonstrated resolution of the order of λ/4. Our analysis in terms of the effective metamaterial parameters indicates that the shape of microlenses and microdroplets provides natural realization of the effective refractive index distribution in the fisheye and inverted Eaton microlenses. The starting point of our analyses is the dispersion law of guided modes in a tapered waveguide. In case of the metal-coated dielectric waveguide it can be written in a simple analytical form:

$$\frac{\omega^2 n_d^2}{c^2} = k_x^2 + k_y^2 + \frac{\pi^2 l^2}{d(r)^2} \tag{14}$$

where $n_d$ is the refractive index of the dielectric, $d(r)$ is the waveguide thickness, and $l$ is the transverse mode number. We assume that the thickness $d$ of the waveguide in the $z$-direction changes adiabatically with radius $r$. If we wish to emulate refractive index distribution $n(r)$ of either 2D fisheye or 2D inverted Eaton lens:

$$\frac{\omega^2 n^2(r)}{c^2} = k_x^2 + k_y^2 \qquad (15)$$

we need to produce the following profile of the microdroplet:

$$d = \frac{l\lambda}{2\sqrt{n_d^2 - n^2(r)}} \qquad (16)$$

This is easy to do for some particular mode $l$ of the waveguide. Typical microdroplet/microlens profiles which emulate the fisheye lens described by equation:

$$n = 2n_1\left(1 + \frac{r^2}{R^2}\right)^{-1} \qquad (17)$$

(where $2n_1$ is the refractive index at the center of the lens, and $R$ is the scale) or the inverted Eaton lens [21] described by:

$$n = 1 \text{ for } r < R, \text{ and } \qquad n = \sqrt{\frac{2R}{r} - 1} \qquad \text{for } r > R. \qquad (18)$$

are shown in Figure 8. Real glycerin microdroplets have shapes, which are somewhere in between these cases. Since the refractive index distribution in the fisheye lens is obtained via the stereographic projection of a sphere onto a plane [7], points near the droplet edge correspond to points located near the equator of the sphere. Therefore, these points are imaged into points located near the opposite droplet edge, as shown in Figure 9(a). The inverted Eaton lens has similar imaging properties, as shown in Figure 9(c). Each droplet depicted in Figure 9 was simulated using scattered field finite element formulation. The continuity of the tangential field components was enforced at the host-droplet interface. The host with the droplet was surrounded by a perfectly matched (absorbing) layer to suppress reflection from the exterior boundaries of the simulation domain.

**Figure 8.** Typical profiles of a microdroplet which emulates either the fisheye lens ($R = 7$ μm) or the inverted Eaton lens ($R = 5$ μm) for the following set of parameters: $l = 1$, $\lambda = 1.5$ μm, $n_d = 1.5$, and $n_1 = 0.65$.

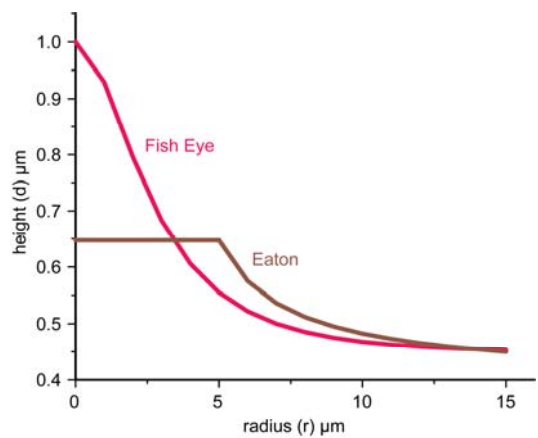

**Figure 9.** Numerical simulations of imaging properties of the fisheye (**a**) and inverted Eaton (**c**) lenses. Points near the edge of the fisheye and Eaton lenses are imaged into opposite points. Refractive index distributions in these lenses are shown to their right in panels (**b**) and (**d**).

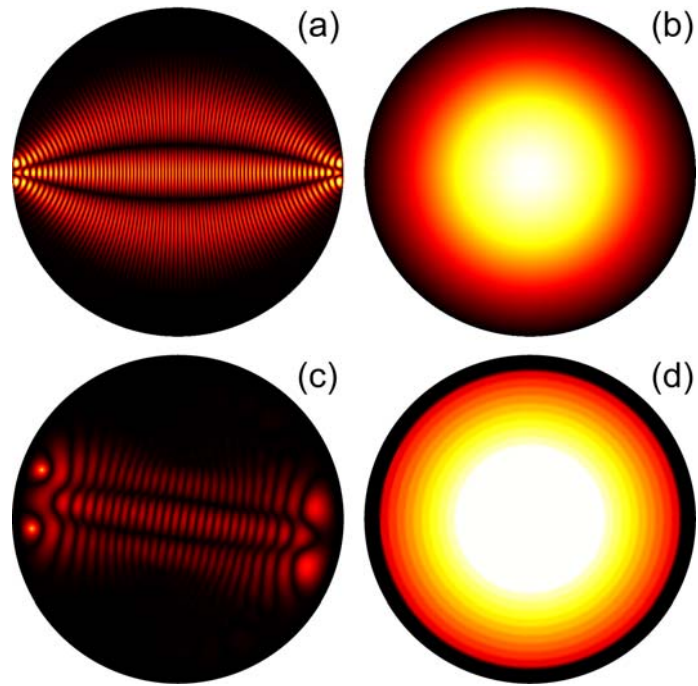

**Figure 10.** Experimental testing of the imaging mechanism of the glycerin microdroplets shown at different magnifications. The droplet is illuminated near the edge with a tapered fiber tip of a near-field scanning optical microscope (NSOM). Image of the NSOM tip is clearly seen at the opposite edge of the droplet.

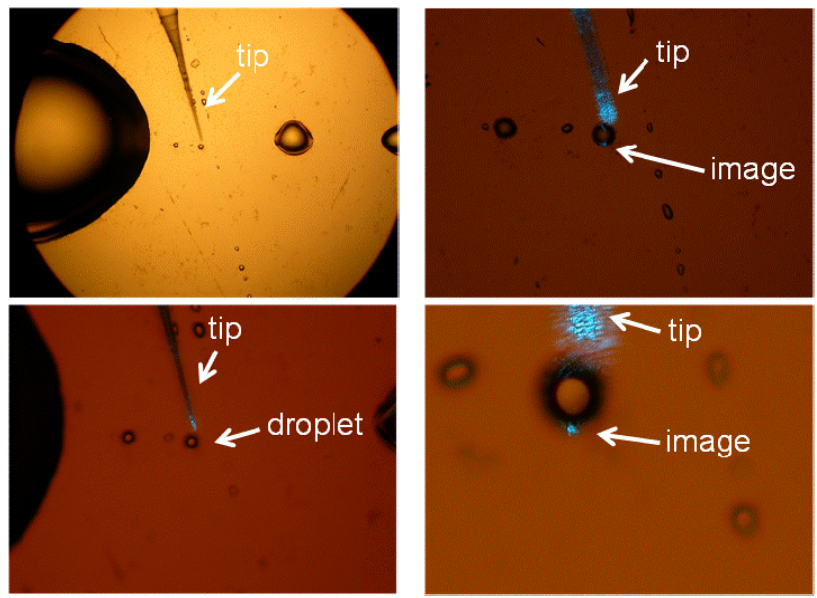

**Figure 11.** Numerical simulations of image magnification (M = 2) using the inverted Eaton lens. Since the sides of the lens play no role in imaging, the overall shape of the imaging device can be altered to achieve the shape of a "deformed droplet".

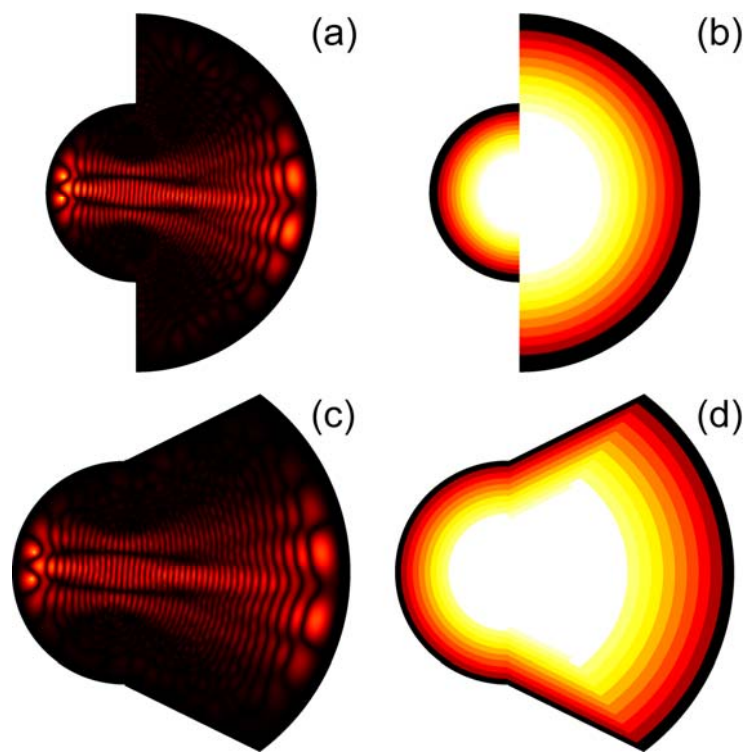

We have tested this imaging mechanism using glycerin microdroplets formed on the surface of gold film, which were illuminated near the edge using tapered fiber tips of a near-field scanning optical microscope (NSOM), as shown in Figure 10. As expected from the numerical simulations, an image of the NSOM tip was easy to observe at the opposite edge of the microdroplet. As has been demonstrated in refs. [22–25], perfect imaging using a Maxwell fisheye or an Eaton lens requires a drain. In our case, the droplet boundary may perhaps act as such a drain. However, more detailed experimental study of this issue is needed.

While the fisheye lens design is difficult to modify to achieve image magnification, modification of the Eaton lens is straightforward. As shown in Figure 11, two halves of the Eaton lens having different values of parameter R can be brought together to achieve image magnification. The image magnification in this case is $M = R_1/R_2$. Our numerical simulations in the case of $M = 2$ are presented. Since the sides of the lens play no role in imaging, the overall shape of the imaging device can be altered to achieve the shape of a "deformed droplet". Using experimental technique described below, we have created glycerin droplets with shapes, which are very close to the shape of the "deformed droplet" used in the numerical simulations. Image magnification of the "deformed droplet" has been tested by moving the NSOM probe tip along the droplet edge, as shown in Figure 12. It appears to be close to the $M = 2$ value predicted by the simulations.

**Figure 12.** Experimental testing of image magnification of the "deformed droplet". The NSOM probe tip was moved along the droplet edge. Bottom row presents results of our numerical simulations in the case of one and two point sources. The shape of the "deformed droplet" used in numerical simulations closely resembles the shape of the actual droplet.

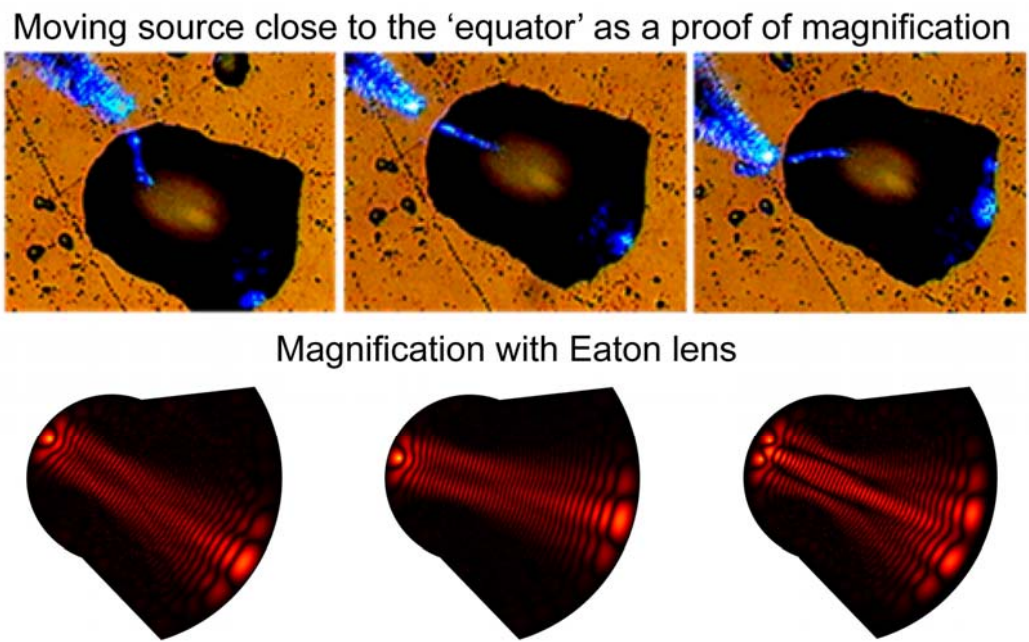

In our imaging experiments the "deformed droplets" were formed in desired locations by bringing a small probe Figure 13(a) wetted in glycerin into close proximity to the sample surface. The probe was prepared from a tapered optical fiber, which has an epoxy microdroplet near its apex. Bringing the probe to a surface region covered with glycerin led to a glycerin microdroplet formation under the probe (Figure 13b). The shape of the glycerin droplet was determined by the shape of the seed droplet of epoxy. Our droplet deposition procedure allowed us to form droplet shapes, which were reasonably close to the shape of a magnifying Eaton lens, as shown in Figures 12 and 14. In addition, the liquid droplet boundary may be expected to be rather smooth because of the surface tension, which is essential for the proper performance of the droplet boundary as a 2D fisheye or Eaton lens.

Image magnification of the 2D magnifying Eaton lens has been measured as demonstrated in Figure 14. Position of the NSOM tip and its image in the second frame is shown by red dots in the first frame. The ratio of the gray line lengths, which connect NSOM tip and image locations in the two frames shown is close to the theoretically predicted value $M = 2$. Thus, we have demonstrated that small dielectric microlenses behave as two-dimensional imaging devices, which can be approximated by 2D fisheye or inverted Eaton lenses. Deformed microlenses/microdroplets were observed to exhibit image magnification, which is consistent with numerical predictions.

**Figure 13.** "Deformed" glycerin droplets were formed in desired locations by bringing a small probe **(a)** wetted in glycerin into close proximity to a sample. The probe was prepared from a tapered optical fiber, which has an epoxy microdroplet near its apex. Bringing the probe to a surface region covered with glycerin led to a glycerin microdroplet formation **(b)** under the probe.

**(a)**                                              **(b)**

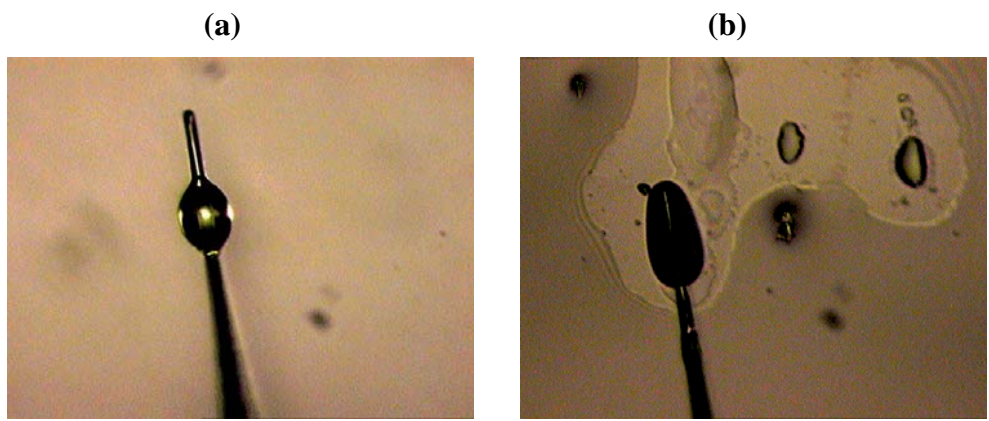

**Figure 14.** Measurements of image magnification by the "deformed droplet": position of the NSOM tip and its image in the second frame is shown by red dots in the first frame.

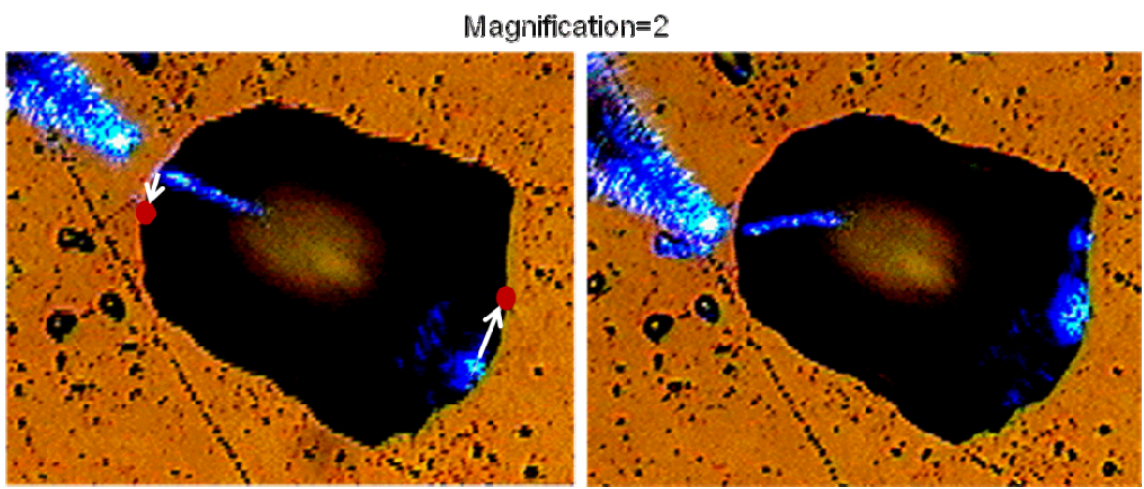

## 5. Conclusions

We have demonstrated that metamaterial devices requiring anisotropic dielectric permittivity and magnetic permeability may be emulated by specially designed tapered waveguides. This approach leads to low-loss, broadband performance, which is difficult to achieve by other means. It is important to note that filling an initial domain between rotationally symmetric curvilinear boundaries, for example, with an anisotropic dielectric allows for independent control over the effective magnetic and electric properties in the equivalent right-cylinder domain. Based on this technique, we have demonstrated such broadband transformation optics devices as electromagnetic cloaks in the visible frequency range operating on a scale ~100 times larger than the wavelength, "trapped rainbow" devices which can be extremely useful in the "spectroscopy on a chip" applications, and novel Maxwell fisheye and Eaton lenses.

**Acknowledgements**

V.N.S. acknowledges support of this research by the NSF grant DMR-0348939; V.M.S. and A.V.K. acknowledge support by ARO-MURI awards 50342-PH-MUR and W911NF-09-1-0539.

**References**

1. Pendry, J.B. Negative refraction makes a perfect lens. *Phys. Rev. Lett.* **2000**, *85*, 3966.
2. Pendry, J.B.; Schurig, D.; Smith, D.R. Controlling electromagnetic fields. *Science* **2006**, *312*, 1780-1782.
3. Leonhardt, U. Optical conformal mapping. *Science* **2006**, *312*, 1777-1780.
4. Smolyaninov, I.I.; Smolyaninova, V.N.; Kildishev, A.V.; Shalaev, V.M. Anisotropic metamaterials emulated by tapered waveguides: application to electromagnetic cloaking. *Phys. Rev. Lett.* **2009**, *102*, 213901.
5. Smolyaninova, V.N.; Smolyaninov, I.I.; Kildishev, A.V.; Shalaev, V.M. Experimental observation of the trapped rainbow. *Appl. Phys. Lett.* **2010**, *96*, 211121.
6. Smolyaninova, V.N.; Smolyaninov, I.I.; Kildishev, A.V.; Shalaev, V.M. Maxwell fisheye and inverted Eaton lenses emulated by microdroplets. *Opt. Lett.* **2010**, *35*, 3396-3398.
7. Leonhardt, U. Perfect imaging without negative refraction. *New J. Phys.* **2009**, *11*, 093040.
8. Schurig, D.; Mock, J.J.; Justice, B.J.; Cummer, S.A.; Pendry, J.B.; Starr, A.F.; Smith, D.R. Metamaterial electromagnetic cloak at microwave frequencies. *Science* **2006**, *314*, 977-980.
9. Smolyaninov, I.I.; Hung, Y.J.; Davis, C.C. Two-dimensional metamaterial structure exhibiting reduced visibility at 500 nm. *Opt. Lett.* **2008**, *33*, 1342-1344.
10. Jacob, Z.; Narimanov, E.E. Semiclassical description of non magnetic cloaking. *Opt. Express* **2008**, *16*, 4597-4604.
11. Landau, L.D.; Lifshitz, E.M. *Quantum Mechanics*; Reed: Oxford, UK, 1988; p. 230.
12. Zayats, A.V.; Smolyaninov, I.I.; Maradudin, A. Nano-optics of surface plasmon-polaritons. *Phys. Rep.* **2005**, *408*, 131-314.
13. Tsakmakidis, K.L.; Boardman, A.D.; Hess, O. "Trapped rainbow" storage of light in metamaterials. *Nature* **2007**, *450*, 397-401.
14. Stockman, M.I. Nanofocusing of optical energy in tapered plasmonic waveguides. *Phys. Rev. Lett.* **2004**, *93*, 137404.
15. Gan, Q.; Ding, Y.J.; Bartoli, F.J. "Rainbow" trapping and releasing at telecommunication wavelengths. *Phys. Rev. Lett.* **2009**, *102*, 056801.
16. Hau, L.V.; Harris, S.E.; Dutton, Z.; Behroozi, C.H. Light speed reduction to 17 metres per second in an ultracold atomic gas. *Nature* **1999**, *397*, 594-598.
17. Vlasov, Y.A.; O'Boyle, M.; Hamann, H.F.; McNab, S.J. Active control of slow light on a chip with photonic crystal waveguides. *Nature* **2005**, *438*, 65-69.
18. Smolyaninov, I.I.; Elliott, J.; Zayats, A.V.; Davis, C.C. Far-field optical microscopy with nanometer-scale resolution based on the in-plane image magnification by surface plasmon polaritons. *Phys. Rev. Lett.* **2005**, *94*, 057401.

19. Lee, J.Y.; Hong, B.H.; Kim, W.Y.; Min, S.K.; Kim, Y.; Jouravlev, M.V.; Bose, R.; Kim, K.S.; Hwang, I.; Kaufman, L.J.; Wong, C.W.; Kim, P.; Kim, K.S. Near-field focusing and magnification through self-assembled nanoscale spherical lenses. *Nature* **2009**, *460*, 498-501.

20. Mason, D.R.; Jouravlev, M.V.; Kim, K.S. Enhanced resolution beyond the Abbe diffraction limit with wavelength-scale solid immersion lenses. *Opt. Lett.* **2010**, *35*, 2007-2009.

21. Minano, J.C. Perfect imaging in a homogeneous three dimensional region. *Opt. Express* **2006**, *14*, 9627-9635.

22. Blaikie, R.J. Comment on 'Perfect imaging without negative refraction'. *New J. Phys.* **2010**, *12*, 058001.

23. Leonhardt, U. Reply to comment on 'Perfect imaging without negative refraction'. *New J. Phys.* **2010**, *12*, 058002.

24. Ma, Y.G.; Ong, C.K.; Sahebdivan, S.; Tyc, T.; Leonhardt, U. Perfect imaging without negative refraction for microwaves. **2010**, arXiv:1007.2530v1.

25. de Rosny J.; Fink, M. Overcoming the diffraction limit in wave physics using a time-reversal mirror and a novel acoustic sink. *Phys. Rev. Lett.* **2002**, *89*, 124301.